\documentclass[reprint,amsmath,amssymb,aps,prb,superscriptaddress]{revtex4-2}
\usepackage{graphicx}% Include figure files
\usepackage{dcolumn}% Align table columns on decimal point
\usepackage{bm}% bold math
\usepackage[hidelinks]{hyperref}% add hypertext capabilities
\usepackage{multirow}
\usepackage{xcolor}
\usepackage{amsmath}
\usepackage{amsfonts}
\usepackage{amssymb}
\usepackage{graphicx}
\newcommand{\BOS}{Bi$_2$O$_2$Se}
\newcommand{\CVS}{cm$^2$V$^{-1}$s$^{-1}$}

\begin{document} 

\title{Giant Modulation of the Electron Mobility in Semiconductor Bi$_2$O$_2$Se via Incipient Ferroelectric Phase Transition} 

\author{Ziye Zhu}
\affiliation{Key Laboratory of 3D Micro/Nano Fabrication and Characterization of Zhejiang Province, School of Engineering, Westlake University, Hangzhou 310024, China}
\affiliation{Institute of Advanced Technology, Westlake Institute for Advanced Study, Hangzhou 310024, China}
\affiliation{School of Materials Science and Engineering, Zhejiang University, Hangzhou 310027, China}

\author{Xiaoping Yao}
\affiliation{Key Laboratory of 3D Micro/Nano Fabrication and Characterization of Zhejiang Province, School of Engineering, Westlake University, Hangzhou 310024, China}
\affiliation{Institute of Advanced Technology, Westlake Institute for Advanced Study, Hangzhou 310024, China}
\affiliation{School of Materials Science and Engineering, Zhejiang University, Hangzhou 310027, China}

\author{Shu Zhao}
\affiliation{Key Laboratory of 3D Micro/Nano Fabrication and Characterization of Zhejiang Province, School of Engineering, Westlake University, Hangzhou 310024, China}
\affiliation{Institute of Advanced Technology, Westlake Institute for Advanced Study, Hangzhou 310024, China}
\affiliation{School of Materials Science and Engineering, Zhejiang University, Hangzhou 310027, China}

\author{Xiao Lin}
\affiliation{Key Laboratory for Quantum Materials of Zhejiang Province, School of Science, Westlake University, Hangzhou 310024, China}
\affiliation{Institute of Natural Sciences, Westlake Institute for Advanced Study, Hangzhou 310024, China}

\author{Wenbin Li}
\email{liwenbin@westlake.edu.cn}
\affiliation{Key Laboratory of 3D Micro/Nano Fabrication and Characterization of Zhejiang Province, School of Engineering, Westlake University, Hangzhou 310024, China}
\affiliation{Institute of Advanced Technology, Westlake Institute for Advanced Study, Hangzhou 310024, China}

% Include the date command, but leave its argument blank.

\date{\today}

%%%%%%%%%%%%%%%%% END OF PREAMBLE %%%%%%%%%%%%%%%%

\begin{abstract}
High-mobility layered semiconductors have the potential to enable the next-generation electronics and computing. This paper demonstrates that the ultrahigh electron mobility observed in the layered semiconductor Bi$_2$O$_2$Se originates from an incipient ferroelectric transition that endows the material with a robust protection against mobility degradation by Coulomb scattering. Based on first-principles calculations of electron-phonon interaction and ionized impurity scattering, it is shown that the electron mobility of Bi$_2$O$_2$Se can reach 10$^4$ to 10$^6$ cm$^2$V$^{-1}$s$^{-1}$ over a wide range of realistic doping concentrations. Furthermore, a small elastic strain of 1.7\% can drive the material toward a unique interlayer ferroelectric transition, resulting in a large increase in the dielectric permittivity and a giant enhancement of the low-temperature electron mobility by more than an order of magnitude. These results establish a new route to realize high-mobility layered semiconductors via phase and dielectric engineering.
\end{abstract}

\maketitle 

\section{Introduction}

Layered and two-dimensional (2D) semiconductors receive prominent attention for their prospects of fabricating high-performance electronic and optoelectronic devices with flexibility and stretchability~\cite{Akinwande2014,Chhowalla2016,Liu2021}. Indeed, the realization of conformal, integrated nanoelectronics based on high-mobility 2D semiconductors could revolutionize the next-generation cyber-physical systems~\cite{Serpanos2018}. This prospect, however, has been hindered by the lack of 2D semiconductors that have both high electron mobility \textit{and} high chemical stability.

Recently, bismuth oxyselenide (\BOS) has emerged as an air-stable, high-mobility layered semiconductor with unprecedented optical sensitivity~\cite{Wu2017a, Wu2019, Wu2017b, Chen2018, Yin2018}. Experimental work has demonstrated a high electron Hall mobility ($>$20,000 \CVS) and Shubnikov-de Haas quantum oscillations in ultrathin films of Bi$_2$O$_2$Se at low temperatures~\cite{Wu2017a}. It is remarkable that such \BOS\ films, grown by chemical vapor deposition (CVD) and having a thickness around 10 layers ($\sim$6 nm), possess a low-temperature electron mobility that rivals those observed in CVD-grown graphene~\cite{Petrone2012} and those at the LaAlO$_3$-SrTiO$_3$ interface~\cite{Ohtomo2004}. The measured room-temperature mobility, averaged at around 200 \CVS~\cite{Wu2017a, Wu2019}, is also rather high for an oxychalcogenide material that shares more chemical similarity with oxides than chalcogenides. Furthermore, in dilute metallic samples of \BOS, $T$-square resistivity without Umklapp scattering was observed at low temperatures, whose microscopic origin remains unclear~\cite{Wang2020}. These intriguing experimental observations call for a comprehensive understanding of the carrier transport properties of \BOS.

In this work, we combine state-of-the-art \textit{ab initio} calculations of phonon-limited intrinsic mobility of \BOS\ with first-principles informed analytical modeling of charged impurity scattering, to offer an in-depth investigation of the electron transport properties of \BOS.  Surprisingly, we find that the high electron mobility of \BOS\ is intricately connected to an incipient interlayer ferroelectric phase transition, which endows \BOS\ with a large low-frequency relative permittivity (static dielectric constant), $\epsilon_0=195.5$, more than an order of magnitude larger than other layered semiconductors such as MoS$_2$. The huge static dielectric permittivity strongly suppresses extrinsic Coulomb scattering such as ionized impurity scattering, allowing \BOS\ to possess a low-temperature electron mobility as large as $10^4$ to $10^6$ \CVS\ over a wide range of realistic doping levels. Such strong dielectric screening also provides \BOS\ a robust protection against mobility degradation by external Coulomb scatterings at room temperature, making the carrier mobility insensitive to Coulombic impurities for doping concentrations up to $10^{19}$~cm$^{-3}$.

More intriguing, we discover that the static dielectric constant of \BOS\ can be further increased dramatically ($\epsilon_0$ reaching several thousand and above), by driving the material closer to the ferroelectric transition. This can be realized by imposing a small biaxial elastic strain of less than 1.7\%, which could be achieved experimentally through epitaxial growth of \BOS\ on perovskite oxide substrates. Such intrinsic dielectric engineering, harnessing the divergent change of dielectric permittivity near a ferroelectric phase transition, allows the increase of electron mobility at low temperature by orders of magnitude and further protects the electron mobility at room temperature. Crucially, we find that the phenomenon of significant mobility enhancement due to strain-induced incipient ferroelectricity is not limited to \BOS, but could be present in many chalcogenide semiconductors in the vicinity of a ferroelectric phase transition.

\section{Results}

%%%%%%%%%%%%%%%%%% Figure 1 %%%%%%%%%%%%%%%%%%%%%
\begin{figure*}[t!]
	\centering
	\includegraphics[width=1.0\textwidth]{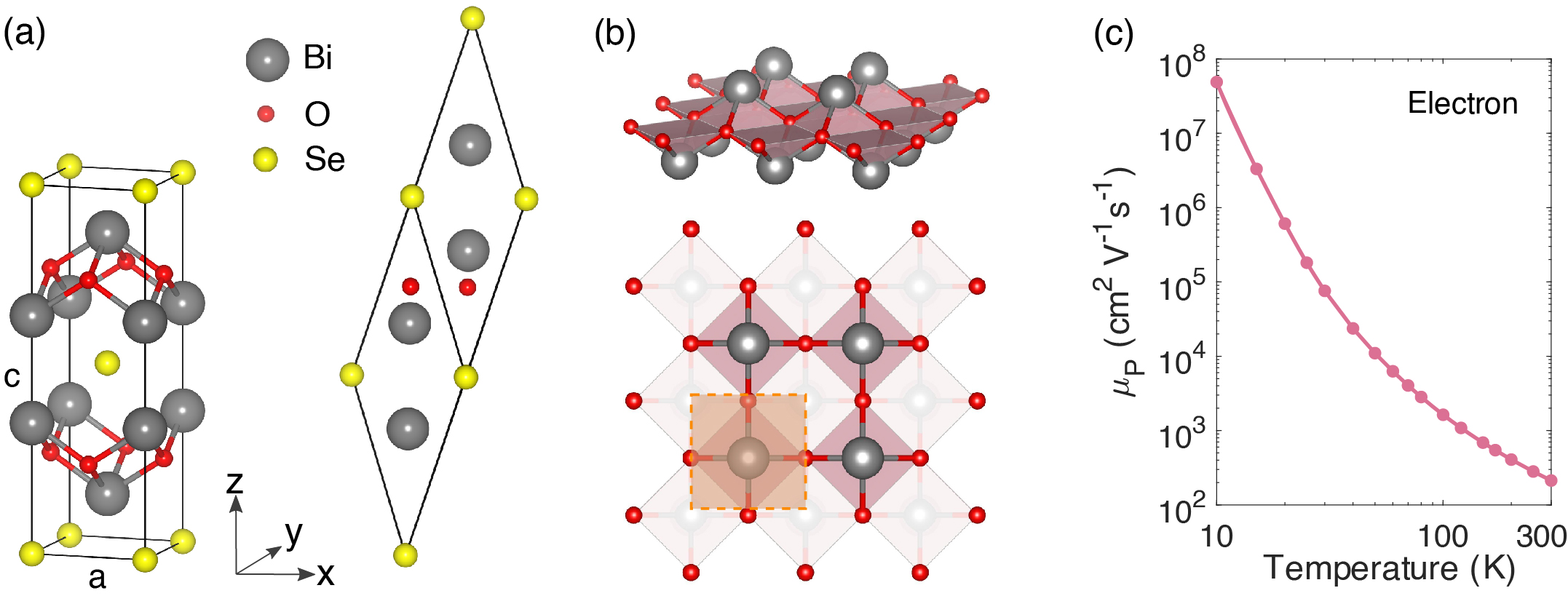}
	\caption{\textbf{Atomistic models and \textit{ab initio} intrinsic electron mobilities of \BOS.} (\textbf{a}) Tetragonal conventional unit cell and rhombohedral primitive cell of Bi$_2$O$_2$Se. (\textbf{b}) Side and top views of the Bi$_2$O$_2$ layers. The shaded orange box outlines the square base of the tetragonal unit cell. (\textbf{c}) Intrinsic phonon-limited electron mobilities of \BOS\ as a function of temperature in the in-plane direction, calculated using the \textit{ab initio} Boltzmann transport equation.}
	\label{fig:fig1}
\end{figure*}

Bi$_2$O$_2$Se crystallizes into a tetragonal \textit{anti}-ThCr$_2$Si$_2$ type of structure~\cite{Boller1973,Hoffmann1985}, with $I4/mmm$ space group symmetry ($a$ = 3.89 \AA, $c$ = 12.21 \AA, see Figure~1a). The Bi$_2$O$_2$ layers, in which the Bi atoms reside at the apex of a square pyramid of O atoms (Figure~1b), are sandwiched between planar Se layers arranged in square lattices. Individual Se atoms have eight nearest-neighbour Bi atoms, which form the eight vertices of a square prism. The height of the square prism, corresponding to the shortest distance of Bi atoms on adjacent Bi$_2$O$_2$ layers, is smaller than the base edge length (3.54 \AA\ versus 3.89 \AA). Charge transfer occurs in the direction from the Bi$_2$O$_2$ layers to Se layers, and the interaction between Bi$_2$O$_2$ and Se layers is considered to be weakly electrostatic. The strength of the interaction between the Bi$_2$O$_2$ and Se layers is much weaker than that of the covalent interaction within individual Bi$_2$O$_2$ layers. As a result, the cleavage of \BOS\ is typically along the Se planes, with the cleavage process leaving 50\% of Se atoms in each Bi$_2$O$_2$ plane~\cite{Chen2018}. A conventional body-centered tetragonal cell consisting of two \BOS\ formula units, or a rhombohedral primitive cell of a single formula unit, can be chosen to represent the crystal, as illustrated in Figure~1a. 

%%%%%%%%%%%%%%%%%%%%%%%  Figure 2 %%%%%%%%%%%%%%%%%%%%%%%%%%
\begin{figure*}[t!]
	\centering
	\includegraphics[width=1.0\textwidth]{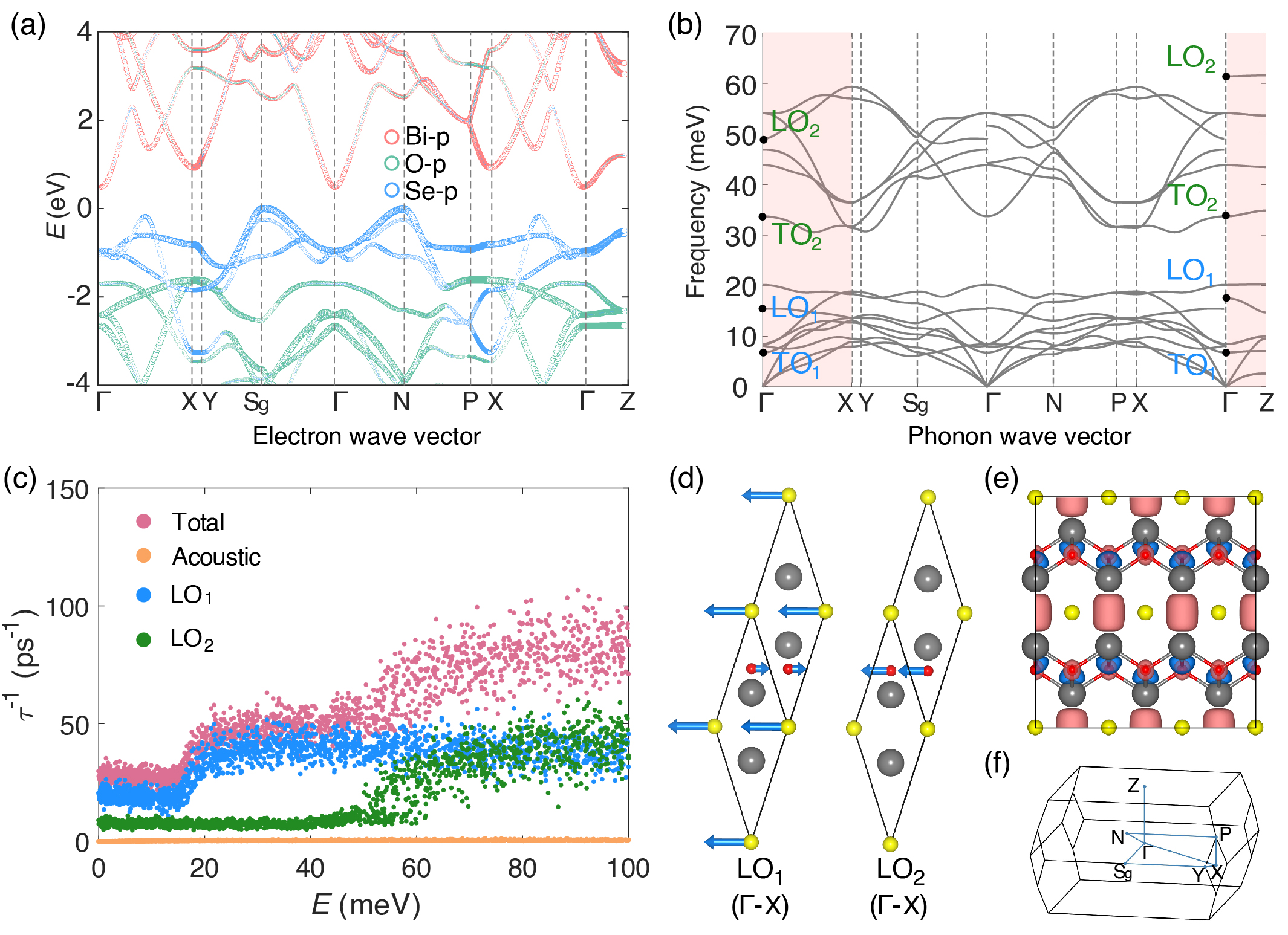}
	\caption{\textbf{Electrons, phonons, and electron-phonon interaction in \BOS.} (\textbf{a}) Electronic band structure of \BOS\ calculated from density functional theory (DFT). The atomic orbital characters of the bands are indicated using open circles of different colors. (\textbf{b}) Phonon spectrum of \BOS\ computed using density functional perturbation theory (DFPT). Two branches of polar longitudinal optical (LO) phonons are indicated by LO$_1$ and LO$_2$, respectively. Their corresponding transverse optical (TO) phonons are represented by TO$_1$ and TO$_2$, respectively. (\textbf{c}) Mode-resolved electron-phonon scattering rates of the conducting electrons as a function of the carrier energy at 300~K. The energy zero is set at the conduction band minimum (CBM). The magenta dots represent the total electron-phonon scattering rates at different carrier energies, whereas the orange, blue, and green dots represent the contributions to the total electron scattering rates from the interactions with acoustic, LO$_1$, and LO$_2$ phonons, respectively.   (\textbf{d}) Atomic displacement patterns corresponding to the two polar LO phonon modes along the $\Gamma$-X direction. (\textbf{e}) Isosurfaces of the CBM electron wave function. (\textbf{f}) Illustration of the reciprocal space path employed in the calculations of electron and phonon band structures in (\textbf{a}) and (\textbf{b}).}
	\label{fig:fig2}
\end{figure*}

We obtain the intrinsic, phonon-limited electron mobilities of \BOS\ using the \textit{ab initio} Boltzmann transport equation (BTE) formalism that treats the electron-phonon interaction (EPI) at a fully first-principles level (see Method in Supporting Information)~\cite{Ponce2020, Giustino2017, Giustino2007}. The calculated intrinsic electron mobilities, shown in Figure~1c, demonstrate an in-plane electron mobility of $\mu_{\text{P}} = 212$ \CVS\ at room temperature, in excellent agreement with the experimental value of $\sim$200~\CVS~\cite{Wu2017a, Wu2019}. The intrinsic mobility increases rapidly with the decrease of temperature, and at 10~K, a huge phonon-limited mobility value of $4.9\times 10^7$ \CVS\ is reached.

To understand the ultrahigh intrinsic mobilities, we first present in Figure~2a the electronic band structure of \BOS\ as calculated from density functional theory (DFT). The associated reciprocal space path used in the band-structure plot is illustrated in Figure~2f. The conduction band minimum (CBM) of \BOS\ is located at the $\Gamma$ point of the Brillouin zone~\cite{Wu2017a}. The calculated in-plane and out-of-plane electron effective masses, $0.13m_0$ and $0.40m_0$ ($m_0$ is the free-electron mass), respectively, are in good agreement with the corresponding experimental values of $0.14m_0$~\cite{Wu2017a, Wang2020} and $0.37m_0$~\cite{Xu2021,Wang2020}. The conduction band of \BOS\ close to the CBM was found to mainly derive from the $p_z$ orbitals of Bi and, to a lesser extent, the $s$ orbitals of O. On the other hand, the valence bands close to the band edge have a predominately Se $p_x/p_y$ character (Figure~S1 and~S2). The orbital character of the CBM state can be further identified from the isosurfaces of the wave function, as shown in Figure~2e, where a strong hybridization between the Bi $p_z$ orbitals on neighbouring Bi$_2$O$_2$ layers can be clearly observed. Hence, despite the quasi-layered structure of \BOS, the conducting electronic states near the CBM are three-dimensional.

The calculated phonon spectrum of \BOS, shown in Figure~2b, is notable for the existence of several low-lying optical phonon modes with energies below 10 meV, as well as the presence of sizable frequency splittings between the polar longitudinal optical (LO) phonons and their corresponding transverse optical (TO) modes. Two sets of polar LO phonons can be identified from the calculated phonon displacement patterns. For phonon wavevectors in the in-plane $\Gamma$--X direction, the lower-lying polar LO phonon branch (hereafter referred to as the LO$_1$ mode) has an energy of 15~meV at the $\Gamma$ point. Its corresponding TO$_1$ branch, which is the lowest-energy optical phonon in \BOS, has a vibrational energy of 7~meV. The higher-lying polar LO$_2$ mode has a significantly higher vibrational energy of 49~meV, and the energy of its corresponding TO$_2$ mode is 34 meV.

The calculated \textit{ab~initio} mode-resolved scattering rates of the electron carriers at 300~K, shown in Figure~2c, demonstrate that the polar LO$_1$ and LO$_2$ modes, which generate Fr\"ohlich interaction with the carriers~\cite{Frohlich1954}, dominate the electron-phonon scattering. In particular, the lower-frequency LO$_1$ mode has the most significant contribution among all phonon modes. The energy dependence of the carrier scattering rates of each polar LO mode is typical of carrier relaxation via Fr\"ohlich interaction~\cite{Li2019, Ponce2019, Lundstrom2000}. The abrupt increases in the scattering rates of LO$_1$ and LO$_2$ modes at $\sim$15 and $\sim$50~meV, respectively, correspond to the onset of carrier relaxation via phonon emission. In general, at room temperature, the scattering rates of the electron carriers whose energies are within 100 meV to the CBM will all have an influence on the carrier mobility. However, if one wishes to analyze the electron mobility in terms of a single electronic state, the most relevant energy will be $3k_B T/2$ above the CBM ($k_B$ is the Boltzmann constant and $T$ is temperature)~\cite{Ponce2019}, whose value is roughly 39~meV at room temperature. At this energy, Figure~2c indicates that the total scattering rate is approximately 50~ps$^{-1}$. Using the simple Drude formula of carrier mobility $\mu = e\tau/m^*$~\cite{Ashcroft1976}, where $\tau$ is the carrier lifetime (the inverse of scattering rate), we would obtain a carrier mobility of 272~\CVS, which is close to our \textit{ab~initio} result of 212~\CVS. If the contributions from carriers of higher energies and relaxation rates are included, the estimated mobility will become closer to the converged \textit{ab~initio} result. 

The calculated eigendisplacements of the two polar LO modes are shown in Figure~2d. The result indicates that the lower-frequency LO$_1$ mode can be interpreted as the ``interlayer polar LO mode": namely, the LO$_1$ mode mostly involves the displacements of the Se layers with respect to the Bi$_2$O$_2$ layers. On the other hand, the higher-frequency LO$_2$ mode can be interpreted as the ``intralayer polar LO mode", which mostly entails the displacements of O atoms with respect to Bi atoms within the Bi$_2$O$_2$ layers. The calculated Fr\"ohlich coupling constants~\cite{MahanBook} of the LO$_1$ and LO$_2$ modes are equal to 0.70 and 0.39, respectively, which indicate a relatively weak electron-phonon coupling. The larger coupling constant of the LO$_1$ mode than the LO$_2$ mode is also consistent with the stronger electron-phonon scattering rates from the LO$_1$ mode in Figure~2c. 

\begin{figure*}[t!]
	\centering
	\includegraphics[width=1.0\textwidth]{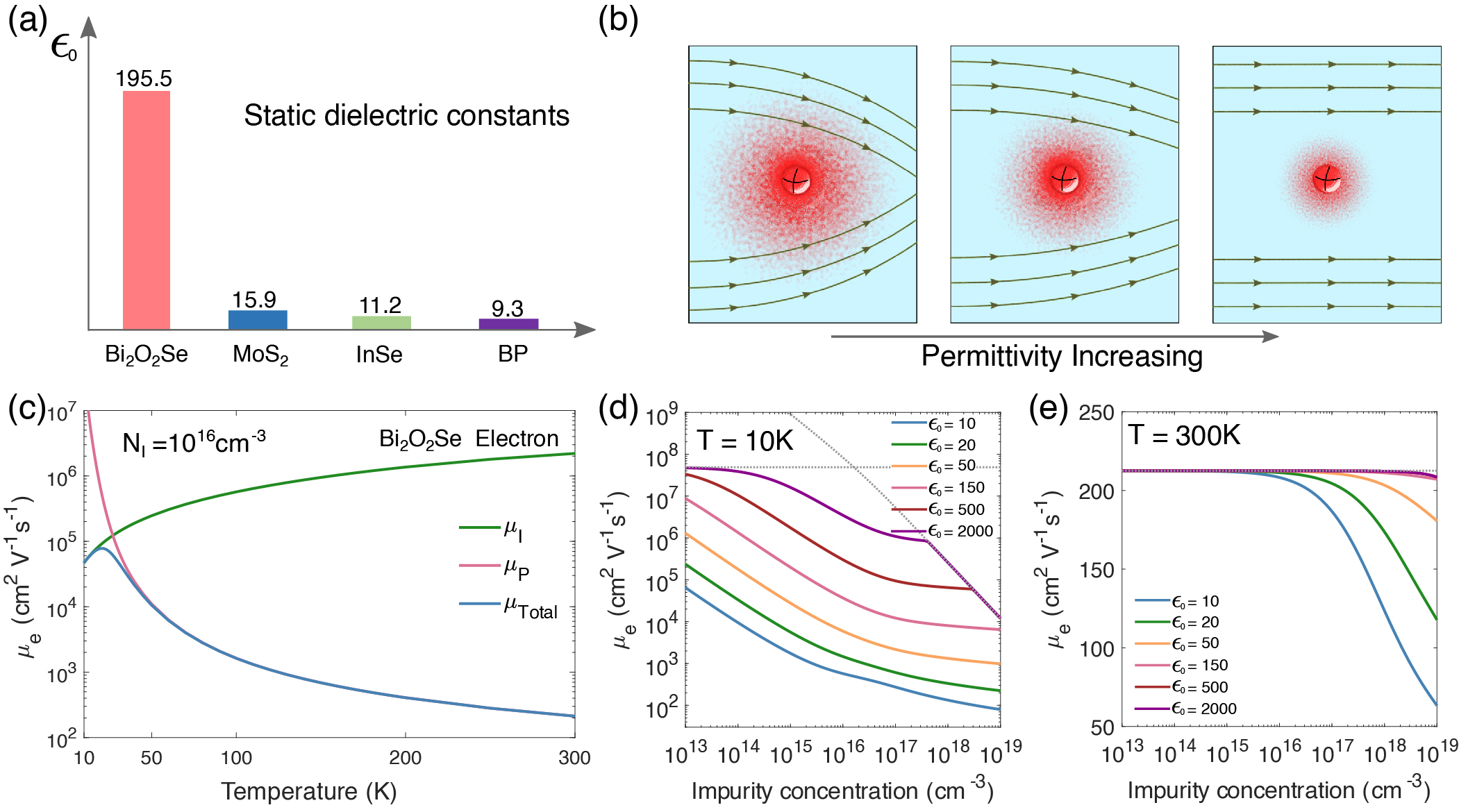}
	\caption{\textbf{Large static dielectric permittivity and its effect on the carrier mobility of \BOS.} (\textbf{a}) Comparison of the static dielectric constant ($\epsilon_0$) of Bi$_2$O$_2$Se with other layered semiconductors, including MoS$_2$, InSe, and black phosphorus (BP). (\textbf{b}) Schematic illustrating the effect of permittivity increase on the scattering of electrons by ionized impurities. The red sphere in the middle represents an ionized impurity with a positive charge, and the lines with arrows represent the moving paths of electrons. Increasing the dielectric permittivity enhances the screening of the Coulomb potentials from ionized impurities, leading to a reduced carrier scattering. (\textbf{c}) Electron mobility of Bi$_2$O$_2$Se as a function of temperature at an ionized impurity concentration of $N_I = 10^{16}$ cm$^{-3}$. $\mu_{\text{I}}$, $\mu_{\text{P}}$, and $\mu_{\text{Total}}$ represent the ionized impurity-limited mobility, phonon-limited mobility, and the total mobility, respectively. $\mu_{\text{Total}}$ were calculated from $\mu_{\text{I}}$ and $\mu_{\text{P}}$ using the Matthiessen's rule: $\mu_{\text{Total}}^{-1} = \mu_{\text{I}}^{-1} + \mu_{\text{P}}^{-1}$.  (\textbf{d},\textbf{e}) $\mu_{\text{Total}}$ calculated for different values of $\epsilon_0$ and $N_I$, at 10~K~(\textbf{d}) and 300~K~(\textbf{e}). Note that in the present model we assume that the phonon-limited mobilities for different values of $\epsilon_0$ are the same and that the carrier concentrations are the same as the corresponding ionized impurity concentrations. The gray horizontal dashed line indicates $\mu_{\text{P}}$. The gray tilted dashed line represents the maximally attainable $\mu_{\text{I}}$, namely, $\mu_{\text{I}}^c$.}
	\label{fig:fig3}
\end{figure*}

At low temperature, Coulomb impurities in the form of ionized dopants or ionized vacancies can play a deciding role in the electron transport of semiconductors. We have therefore investigated the effect of ionized impurity scattering on the carrier mobility of \BOS, by means of the Brooks-Herring model and the BTE~\cite{Brooks1955, Chattopadhyay1981}. Within this model, the scattering potentials of ionized impurities take the form of screened Coulomb potentials: $V(r) = \frac{Ze}{4\pi \varepsilon_0 \epsilon_0 r} \exp(-q_D r)$, where $r$ is the distance to the scattering center, $Ze$ the charge of the impurity, $\varepsilon_0$ the vacuum permittivity, and $\epsilon_0$ the static dielectric constant. $q_D = \sqrt{e^2n^*/\varepsilon_0 \epsilon_0 k_B T}$ represents the reciprocal Debye screening length, in which $n^*$ is the effective screening carrier density~\cite{Wolfe1988}. The Brooks-Herring model introduces an ionized impurity scattering rate that depends on the wavevector $k$ of the electron:
\begin{equation}
	\frac{1}{\tau_{\mathrm{BH}}} = \frac{N_I Z^2 e^4 m^*}{8\pi \hbar^3 \varepsilon_0^2 \epsilon_0^2 k^3} \left[\ln (1+b) - b/(1+b) \right].
	\label{eq:BH_scat_rate_main}
\end{equation}
Here, $N_I$ is the ionized impurity concentration. The dimensionless parameter $b$ is given by $b\equiv 4k^2/q_D^2$~\cite{Chattopadhyay1981}. The ionized impurity-limited mobility, $\mu_{\text{I}}$, can be calculated by integrating over all electronic states using the BTE (Supporting Information), leading to
\begin{equation}
	\mu_{\text{I}} = \frac{32 \sqrt{2}\,\pi \varepsilon_0^2 \epsilon_0^2}{3N_I Z^2 e^3 \sqrt{m^*} k_B T} \frac{\int dE\, E^3 f^0(1-f^0) G(b)}{\int dE \, E^{1/2} f^0},
	\label{eq:mob_I_main}
\end{equation}
where $G(b) = \left[\ln(1+b) - b/(1+b)\right]^{-1}$, with $b$ now written in terms of energy as $b =  8m^*E/{\hbar}^2 q_D^2$. $f_0$ is the Fermi-Dirac distribution, and $E$ denotes the carrier energy above the CBM. Importantly, the Coulomb cross section of ionized impurity scattering has a lower bound of $\sigma_c \approx a^2 $, where $a$ is the lattice constant~\cite{Huang2021, Morita1963}. This imposes an upper bound to $\mu_{\text{I}}$ as given by (Supporting Information):
\begin{equation}
	\mu_{\text{I}}^c = \frac{\sqrt{2}e}{3N_I a^2 \sqrt{m^*}k_B T} \frac{\int dE \,E f^0(E)[1 - f^0(E)]}{\int dE \, E^{1/2} f^0(E) }.
\end{equation}
On the basis of Eq.~\ref{eq:mob_I_main}, when the calculated ionized impurity-limited mobility $\mu_{\text{I}}$ is below the upper bound set by $\mu_{\text{I}}^c$, the value of $\mu_{\text{I}}$ increases quadratically with $\epsilon_0$. This indicates that the low-frequency dielectric permittivity can have a pronounced effect on ionized impurity-limited mobility.

Surprisingly, our calculation reveals that \BOS\ possesses a huge in-plane static dielectric constant of $\epsilon_0^{\perp} = 195.5$. This value is more than an order of magnitude higher than other layered semiconductors such as MoS$_2$~\cite{Laturia2018}, InSe~\cite{Li2020a}, and black phosphorus~\cite{Kumar2016}, as illustrated in Figure~3a. The value of $\epsilon_0$ in the out-of-plane direction, $\epsilon_0^{\parallel} = 117.5$, is also significant. Such a large low-frequency relative permittivity, supported by the recent experimentally measured value of ~155 by one of the authors~\cite{Xu2021}, is expected to strongly suppress ionized impurity scattering in \BOS, as schematically illustrated in Figure~3b. Indeed, as shown in Figure~3c, after including the contributions from both ionized impurity-limited mobility $\mu_{\text{I}}$ and phonon-limited mobility $\mu_{\text{P}}$ using Matthiessen's rule $\mu_{\text{Total}}^{-1} = \mu_{\text{I}}^{-1} + \mu_{\text{P}}^{-1}$, the calculated total electron mobility $\mu_{\text{Total}}$ of \BOS\ at 10~K and an impurity concentration of $N_I = 10^{16}$~cm$^{-3}$ can reach an exceptionally high value of 46,000 \CVS. 

To shed more light on the critical role of low-frequency relative permittivity in realizing the ultrahigh electron mobility of \BOS, we have further calculated $\mu_{\text{Total}}$ as a function of $\epsilon_0$ at different $N_I$ and temperatures. The results at 10 and 300~K are shown in Figure~3d and e, respectively (see also Figure~S3 for results at additional temperatures). At 10~K, $\mu_{\text{I}}$ dominates the contribution to $\mu_{\text{Total}}$. This allows a significant enhancement of $\mu_{\text{Total}}$ through an increase in $\epsilon_0$. Indeed, Figure~3d shows that, at $N_I = 10^{16}$ cm$^{-3}$, when $\epsilon_0$ is varied from 10 (a typical value for semiconductors) to 500, the $\mu_{\text{Total}}$ at 10~K increases from a modest value of $\sim$500 \CVS\ to a huge value of $\sim$300,000 \CVS, representing a 3 orders of magnitude increase.

At room temperature, a large static dielectric constant $\epsilon_0$ further provides \BOS\ a robust protection against mobility degradation by ionized impurity scattering, as shown in Figure~3e. When $\epsilon_0$ is above 150, the total electron mobility $\mu_{\text{Total}}$ shows little variation even as the ionized impurity concentration $N_I$ reaches 10$^{19}$~cm$^{-3}$. In stark contrast, when $\epsilon_0 = 10$, $\mu_{\text{Total}}$ decreases from the intrinsic value of 212~\CVS\ to merely 63~\CVS at $N_I = 10^{19}$~cm$^{-3}$. Our results thus provide a fundamental explanation for the recent experimental observation that the room-temperature electron mobility of \BOS\ shows little variation in the residual carrier concentration~\cite{Wu2019}. These results further suggest a promising new route to find novel high-mobility layered semiconductors that are immune or insensitive to mobility degradation by Coulomb impurity scattering, that is, through the search or design of high-mobility materials with a large intrinsic static dielectric permittivity.

\begin{figure*}[t!]
	\centering
	\includegraphics[width=1.0\textwidth]{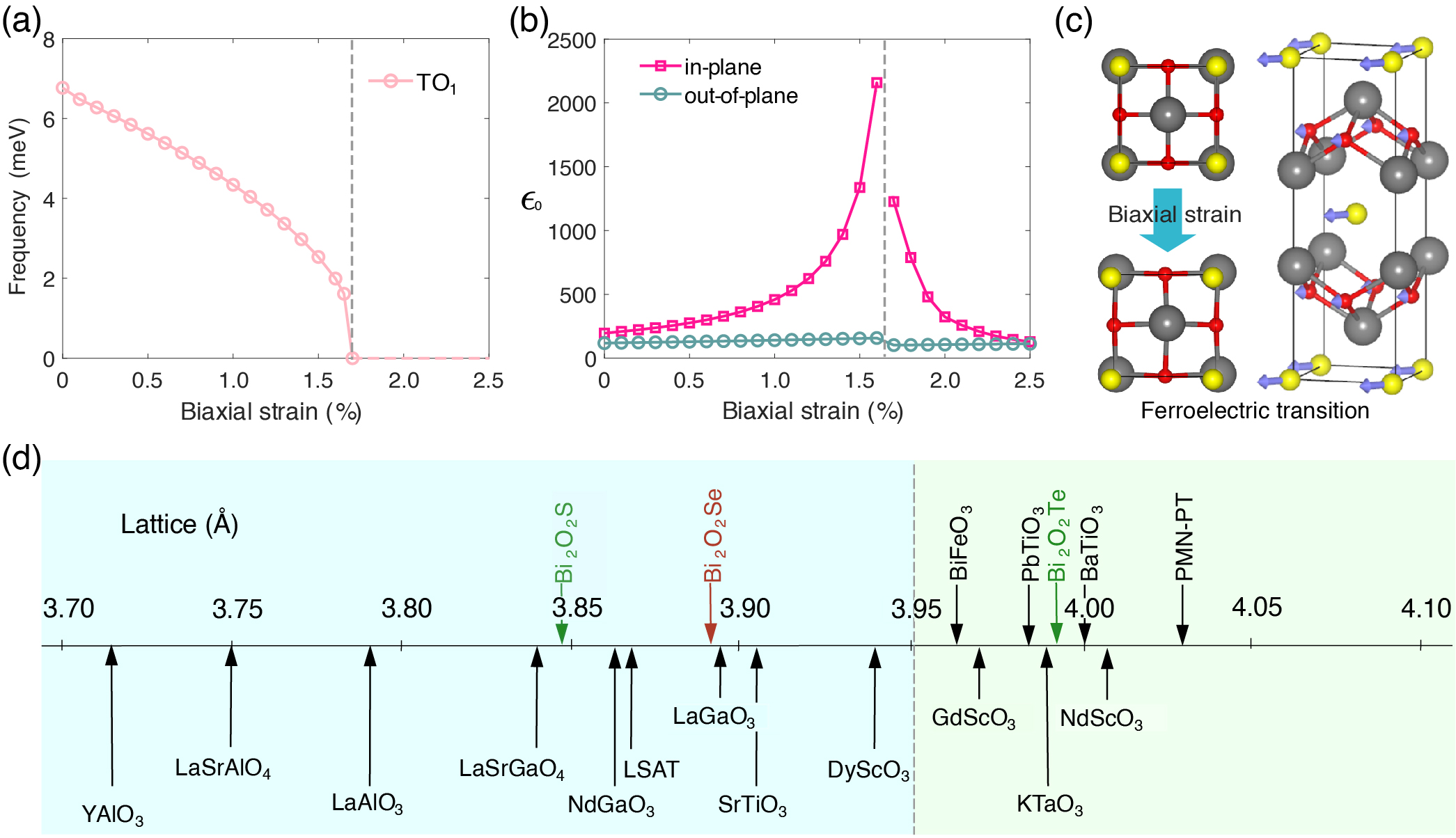}
	\caption{\textbf{Incipient ferroelectricity and giant strain modulation of permittivity in \BOS.} (\textbf{a}) Dependence of the frequency of the lowest-energy TO mode on the magnitude of the in-plane biaxial strain. (\textbf{b}) Strain dependence of the static dielectric constants in both in-plane and out-of-plane directions. (\textbf{c}) Top and side views of the structural distortions involved in the ferroelectric phase transition of Bi$_2$O$_2$Se. The purple arrows superimposed on the atoms represent the directions of atomic displacements during the phase transition. (\textbf{d}) Number line illustrating the in-plane lattice constants (in angstroms) of some perovskites and perovskite-related materials with cubic or (pseudo)tetragonal structure, which could be used as the substrates for epitaxial growth of \BOS\ with coherent interfaces. The lattice constants of Bi$_2$O$_2$X (X = S, Se, Te) are also indicated on the plot. The vertical dashed line located at $\sim$3.95~\AA\ represents the lattice constant at which \BOS\ undergoes the ferroelectric transition. The materials above the number line, except Bi$_2$O$_2$X, are ferroelectric at zero strain and room temperature. All the materials below the number line have commercially available single-crystal substrates. The lattice constant data of the perovskite materials are from ref~\citenum{Schlom2007}.}
	\label{fig:fig4}
\end{figure*}

The large static dielectric constant of \BOS\ was found to originate from the low frequency ($\sim$7~meV) of the infrared-active TO$_1$ mode that corresponds to interlayer shearing between the Se and Bi$_2$O$_2$ layers (Supporting Information). Crucially, we find that the frequency of the TO$_1$ mode can be further reduced by imposing a small biaxial strain, which ultimately drives \BOS\ toward a ferroelectric phase transition. The reduction in the vibrational frequency of the TO$_1$ mode as a function of strain is shown in Figure~4a (see also Figure S4). When the imposed strain reaches 1.7\%, which corresponds to an in-plane lattice parameter of 3.95 \AA, the TO$_1$ mode completely softens. The calculated local potential energy variation with respect to the atomic displacements of the TO$_1$ mode demonstrates that a double-well potential energy surface develops at this critical point (Figure~S5). In the meantime, the calculated static dielectric constant $\epsilon_0^{\perp}$ exhibits a divergence around the critical point (Figure~4b), which is a hallmark of a second-order ferroelectric transition.

The distorted structure of \BOS\ after the ferroelectric transition is shown in Figure~4c. We can see that the transition results in a relative shift of the Se layers with respect to the Bi$_2$O$_2$ layers in the in-plane square diagonal direction. This is accompanied by a slight distortion of the O atoms within the Bi$_2$O$_2$ layers along the same direction, although their displacements are an order of magnitude smaller that those of the Se atoms. The absolute displacements of the Se atoms increase with strain $\xi$ as $\Delta d \propto 1.7 \sqrt{\xi - \xi_c}$~[\AA], where $\xi_c = 0.017$ is the critical strain value. Since the ferroelectric transition mainly involves the displacements of atoms in the Se layers with respect to the Bi$_2$O$_2$ layers, the transition can be regarded as an ``interlayer sliding ferroelectric transition". This strain-induced ferroelectric transition of \BOS\ was previously observed in DFT calculations~\cite{Wu2017Zeng}, but its microscopic mechanism has never been elucidated.

The incipient ferroelectric transition of \BOS\ has profound implications on its carrier transport properties. Indeed, the drastic increase in the static dielectric constant leading up to the transition can be exploited to significantly enhance the carrier mobility of \BOS. Figure~4b shows that the in-plane static dielectric constant of \BOS\ increases from  $\epsilon_0^{\perp} = 195.5$ at zero strain to $\epsilon_0^{\perp} = 458.6$ at a strain equal to 1\%. Further increase of the strain to 1.6\% leads to a huge value of $\epsilon_0^{\perp} = 2158.6$. Meanwhile, the electron effective mass, vibrational energies of the polar LO phonon modes, and the Fr\"ohlich coupling constants, all do not exhibit significant variation (Figure~S6). As a result, the phonon-limited electron mobilities $\mu_{\text{P}}$ of \BOS\ are practically unaffected by these small strains. On the basis of our earlier results, the huge increase in $\epsilon_0$ will result in a drastic enhancement in ionized impurity-limited electron mobility $\mu_{\text{I}}$, which will manifest experimentally as a giant increase in $\mu_{\text{Total}}$ at low temperatures. Indeed, our calculations indicate that, when a 1.6\% of biaxial strain is imposed on \BOS, at an impurity concentration of $10^{16}$~cm$^{-3}$, the total electron mobility of \BOS\ at 10~K leaps from 46,000 \CVS\ to 1,200,000 \CVS, representing an increase of more than an order of magnitude.

The amount of biaxial strain needed to drive \BOS\ toward the ferroelectric transition, $\sim$1.7\%, is eminently achievable experimentally. This is not only because layered semiconductors at the nanoscale can sustain an experimental biaxial strain as large as 10\%~\cite{Bertolazzi2011} but also because it has been experimentally shown that \BOS\ can be epitaxially grown on perovskite oxide substrates such as SrTiO$_3$, LaAlO$_3$, and (La,Sr)(Al,Ta)O$_3$ with perfect lattice matching and strong interaction with the substrates~\cite{Tan2019,Liang2019}. The experimental observations are consistent with the fact that, in \BOS, the interlayer bonding between the Bi$_2$O$_2$ layers and Se layer has an ionic character, making the strength of interlayer interaction significantly stronger in \BOS\ than in other layered materials with van der Waals interaction. Indeed, previous DFT calculations have shown that the binding energy between the Bi$_2$O$_2$ layers and Se layers in \BOS\ is around 77~meV/\AA$^2$ (ref~\citenum{Wei2019}), whereas the interlayer binding energy of MoS$_2$ has a much smaller value of around 20~meV/\AA$^2$ (ref~\citenum{Bjorkman2012}). When \BOS\ is epitaxially grown on oxide substrates, charge transfer may occur at the interface, leading to a strong interfacial binding. Therefore, it is possible to epitaxially strain Bi$_2$O$_2$Se beyond a tensile strain of 0.5\%, which has already been clearly demonstrated on SrTiO$_3$ substrates~\cite{Tan2019,Liang2019}. In Figure~4d, we illustrate the closeness in lattice constants between Bi$_2$O$_2$X (X = S, Se, Te) and many perovskite-related materials~\cite{Schlom2007}. Ternary compounds with the ThCr$_2$Si$_2$ structure type~\cite{Hoffmann1985} that also have similar tetragonal lattice parameters to Bi$_2$O$_2$X are shown in Figure~S7. We conclude that DyScO$_3$ and (Sr,Ba)TiO$_3$ could be suitable substrates for realizing the predicted giant strain-tuning of the dielectric permittivity and carrier mobility of \BOS.

We emphasize that although our study focuses on \BOS, recent experiments have demonstrated that many chalcogenide semiconductors also exhibit ferroelectric transitions near ambient conditions~\cite{BarrazaLopez2021}. In Figure~S8, we show by DFT-calculated results that, in PbS, PbSe, PbTe, and SnTe~\cite{Rabe1985, An2008, Zhang2009, Skelton2014, Chang2016}, strong strain-induced modulations of dielectric constants and potentially electron mobilities could also be realized near the equilibrium conditions.

\section{Discussion}
A giant enhancement of the carrier mobility of a material through an increase of the dielectric constant requires the following two conditions: (1) within a given temperature and carrier concentration range, the carrier mobility of a material is limited by ionized impurity (charged defect) scattering. (2) The ionized impurity scattering can be significantly suppressed by increasing the dielectric constant of the material. Here we have assumed that phonon scattering and ionized impurity scattering are the two major sources of carrier scattering, which is often the case in high-quality semiconductor materials~\cite{YuCardona}. Denoted by $\mu_{\text{P}}$, the phonon-limited mobility, $\mu_{\text{I}}$, the carrier mobility due to the scattering from the screened Coulomb potentials of ionized impurities, and, $\mu_{\text{I}}^c$, the maximally attainable $\mu_{\text{I}}$ at a given ionized impurity concentration (due to the existence of a minimum cross section of ionized impurity scattering that originates from the short-range ``core" of a Coulombic impurity potential~\cite{Morita1963, Huang2021}), the two conditions stated above can be mathematically written as $\mu_{\text{I}} < \mu_{\text{P}}$, $\mu_{\text{I}} < \mu_{\text{I}}^c$. The first condition ($\mu_{\text{I}} < \mu_{\text{P}}$) is usually satisfied at low temperatures in doped semiconductors (due to the suppression of phonon scattering at low temperatures, which leads to a large $\mu_{\text{P}}$) or at any temperature when the ionized impurity concentration is sufficiently high (due to strong ionized impurity scattering, which gives rise to low $\mu_{\text{I}}$). The second condition ($\mu_{\text{I}} < \mu_{\text{I}}^c$) requires that, for the carriers most relevant to charge transport, the average cross section of ionized impurity scattering is still larger than the dimension of the core of the impurity potential, which is on the order of the lattice constant squared~\cite{Morita1963, Huang2021}. Note that even if the first condition $\mu_{\text{I}} < \mu_{\text{P}}$ is not satisfied, as long as $\mu_{\text{I}} < \mu_{\text{I}}^c$ is valid, the total electron mobility $\mu_{\text{Total}}$ can still be improved by suppressing ionized impurity scattering, although in this case, the possible magnitude of increase in $\mu_{\text{Total}}$ will be limited.

Although the first condition $\mu_{\text{I}} < \mu_{\text{P}}$ is more likely to be satisfied in a system of high impurity concentration, the second condition $\mu_{\text{I}} < \mu_{\text{I}}^c$ can be violated if the carrier concentration resulting from the ionization of impurities surpasses a critical value. This is especially the case in uncompensated or weekly compensated systems, where the majority of charge carriers derive from the ionization of either donors or acceptors at or below room temperature. The possible violation of $\mu_{\text{I}} < \mu_{\text{I}}^c$ in a system of high carrier concentration has the following reasons: (i) The Coulombic cross section by an ionized impurity decreases with the increase of the momentum of the incident electron~\cite{Chattopadhyay1981}, which is physically intuitive as the motion of an electron with a higher kinetic energy is less affected by a Coulomb potential. When the carrier concentration $n$ increases beyond a certain limit, the electron gas will become degenerate, and the carriers contributing most to charge conductivity will be those close to the Fermi level. Indeed, in a degenerate semiconductor, the electronic states well below the Fermi level will not contribute to the conductivity~\cite{Ashcroft1976}. In a degenerate electron gas, the Fermi wavevector $k_F$ increases with the electron concentration $n$ as $k_{F} \propto n^{1/3}$. Therefore, when $n$ increases, the corresponding increase of $k_F$ will lead to a rapid decrease of the Coulomb cross section. (ii) The increased free electron concentration will also enhance the screening of the ionized impurities, which contributes to the lowering of the scattering cross section. This is reflected in the decrease of the Debye screening length as the effective screening carrier density increases. As a result of the above two contributions, beyond a critical carrier concentration, the scattering cross section of the electrons near the Fermi level will reach the dimension of the defect core size (on the order of the lattice constant squared) and cannot be decreased further. Correspondingly, the ionized impurity scattering can not be further suppressed and the condition $\mu_{\text{I}} < \mu_{\text{I}}^c$ is no longer valid.

If a semiconductor material has a large static dielectric constant, the additional strong lattice contribution to the screening of Coulomb potentials will significantly lower the threshold of carrier concentration to reach the minimum cross-section limit of ionized impurity scattering. For example, in SrTiO$_3$, which is a quantum paraelectric with a static dielectric constant around 20,000 at 4.2~K~\cite{Mueller1979}, the cross section limit is reached when $n$ is around 10$^{17}$~cm$^{-3}$ (ref~\citenum{Huang2021}). Above this critical carrier concentration, further increase of the dielectric constant would not improve the carrier mobility, since the cross section of ionized impurity scattering relevant to carrier transport cannot be reduced further~\cite{Huang2021}. 

In the transport regime of minimum ionized-impurity scattering cross section, assuming that most of the carriers are from the ionization of impurities ($n \approx N_I)$, the carrier mobility will decrease with the carrier concentration $n$ as $n^{-4/3}$, which was originally derived in refs~\citenum{Morita1963} and~\citenum{Shimizu1963} and experimentally observed at liquid helium temperature in heavily doped PbTe ($n$ above 10$^{18}$~cm$^{-3}$, ref~\citenum{Kanai1961}), moderately to heavily doped SrTiO$_3$ ($n$ above 10$^{17}$~cm$^{-3}$, refs~\citenum{Huang2021},\citenum{Spinelli2010}), and heavily doped KTaO$_3$ ($n$ above $5\times 10^{18}$~cm$^{-3}$, ref~\citenum{Wemple1965}). These materials, including PbTe, SrTiO$_3$ and KTaO$_3$, are all incipient ferroelectric materials with a large static dielectric constant (above 1,000) at low temperatures~\cite{Mueller1979,Wemple1965,Shimizu1979}, therefore the cross-section-limited transport regime can be observed at smaller carrier concentrations than other more conventional semiconductor materials. Given that the crystal and electronic band structures of SrTiO$_3$ and KTaO$_3$ are rather similar~\cite{Usui2010}, the much lower threshold of carrier density to reach the cross-section-limited transport regime in SrTiO$_3$ than that of KTaO$_3$ can be explained by the significantly higher low-temperature static dielectric constant of SrTiO$_3$ (20,000 in SrTiO$_3$ versus 4,400 in KTaO$_3$ at 4.2~K~\cite{Mueller1979,Wemple1965}). The $\mu_{\text{I}}^c$ $\sim$ $n^{-4/3}$ scaling in the cross-section-limited transport regime is also numerically produced in the present work for Bi$_2$O$_2$Se, corresponding to the tilted dashed line in Figure~3d. However, because of the much smaller static dielectric constant of unstrained Bi$_2$O$_2$Se than PbTe, SrTiO$_3$, and KaTaO$_3$, the cross-section-limited transport regime is not reached in Bi$_2$O$_2$Se until $n$ is above $1\times 10^{19}$~cm$^{-3}$.  

The above discussions explain why a giant mobility enhancement through strain-induced incipient ferroelectricity and the resulting increase of dielectric permittivity is possible in Bi$_2$O$_2$Se: (1) Unlike traditional oxide ferroelectrics and incipient ferroelectrics such as BaTiO$_3$ and SrTiO$_3$, respectively, Bi$_2$O$_2$Se is an oxychalcogenide that exhibits a relatively small bandgap ($\sim$0.8~eV~\cite{Chen2018}), small electron effective mass (in-plane effective mass around $0.14m_0$~\cite{Wu2017a}), and a moderate electron-phonon interaction (Fr\"ohlich coupling constant of $\sim$0.7 and $\sim$0.39 for the two polar LO modes). Hence, the intrinsic phonon-limited mobility of Bi$_2$O$_2$Se is much higher than that of SrTiO$_3$ ($\sim$200~cm$^2$V$^{-1}$s$^{-1}$ of Bi$_2$O$_2$Se~\cite{Wu2017a,Wu2019} versus 7~cm$^2$V$^{-1}$s$^{-1}$ of SrTiO$_3$~\cite{Cain2013} at 300~K). The electron mobility of BaTiO$_3$ is even lower, with a value of 0.3~cm$^2$V$^{-1}$s$^{-1}$ at 300~K, and the mobility is still only 12~cm$^2$V$^{-1}$s$^{-1}$when the temperature is reduced to 120~K. This indicates that in BaTiO$_3$ the condition of $\mu_{\text{I}} < \mu_{\text{P}}$ may be difficult to realize, even when the temperature is lowered much further~\cite{Kolodiazhnyi2003}. In contrast, our first-principles calculations have shown that, at 10~K, the phonon-limited mobility $\mu_{\text{P}}$ of Bi$_2$O$_2$Se is as high as $5\times 10^7$~cm$^2$V$^{-1}$s$^{-1}$, whereas the ionized-impurity-limited mobility $\mu_{\text{I}}$ at the same temperature and a carrier concentration of 10$^{16}$~cm$^{-3}$ is around $5\times 10^4$~cm$^2$V$^{-1}$s$^{-1}$, a value that is 3 orders of magnitude lower than $\mu_{\text{P}}$. Given that $\mu_{\text{I}} \ll \mu_{\text{P}}$ is valid over a wide range of temperatures and carrier concentrations in Bi$_2$O$_2$Se, its electron mobility, when limited by $\mu_{\text{I}}$, has significant room for further improvement. (2) Unlike SrTiO$_3$, the static dielectric constant of Bi$_2$O$_2$Se, whose calculated and experimentally measured values are in the range between 150 and 200, without a strong temperature dependence between 2~K and room temperature~\cite{Xu2021}, can still be significantly increased via biaxial strain before the electron transport reaches the cross-section-limited regime. Indeed, as shown in Figure~3d, the low-temperature electron mobility of Bi$_2$O$_2$Se is still in the $\mu_{\text{I}} < \mu_{\text{I}}^c$ regime below a carrier concentration of $3\times 10^{17}$~cm$^{-3}$ when its dielectric constant is increased to 2,000. The possibility to maintain $\mu_{\text{I}} \ll \mu_{\text{P}}$ and $\mu_{\text{I}} \ll \mu_{\text{I}}^c$ over a wide range of temperatures and carrier concentrations in Bi$_2$O$_2$Se, even as its dielectric constant is significantly increased, explains why its calculated low-temperature electron mobility can be increased by more than an order of magnitude through strain-induced enhancement of dielectric permittivity. 

We note that, although a giant strain-induced enhancement of low-temperature mobility through the suppression of ionized impurity scattering has not been reported before, materials with large static dielectric constants, such as PbS/PbSe/PbTe, SrTiO$_3$, and KaTiO$_3$, do show exceptionally high low-temperature electron mobilities~\cite{Spinelli2010, Wemple1965, Cain2013,Allgaier1958,Tufte1967}. Mobility values as high as 800,000~cm$^2$V$^{-1}$s$^{-1}$ at 4.2~K in PbTe~\cite{Allgaier1958}, 22,000-50,000~cm$^2$V$^{-1}$s$^{-1}$ at 2~K in SrTiO$_3$~\cite{Spinelli2010, Cain2013, Tufte1967}, and 23,000~cm$^2$V$^{-1}$s$^{-1}$ at 4.2~K in KTaO$_3$ have been experimentally measured~\cite{Wemple1965}. In addition, in KTa$_{1-x}$Nb$_x$O$_3$ (KTN) crystals, whose paraelectric-ferroelectric transition temperature $T_c$ can be tuned by the composition of Nb, mobility peaks were observed near $T_c$, where the dielectric constant also peaks~\cite{Siemons2012}. The low measured carrier concentrations in the KTN crystals suggest that the Coulombic defect scattering is not in the cross-section-limited regime, even as the static dielectric constants close to $T_c$ are already on the order of 1000. Hence, the mobility of the KTN crystals could still be improved via a further increase of the dielectric constant as the $T_c$ is approached. These results indicate that a further significant increase of the electron mobility of Bi$_2$O$_2$Se through strain-induced enhancement of dielectric permittivity and the resulting suppression of ionized impurity is indeed possible.

We further note that several other recently discovered complex chalcogenide semiconductors, including BaZrS$_3$ and Ba$_3$Zr$_2$S$_7$, also show high static dielectric constants in the range of 50-100~\cite{Filippone2020, Jaramillo2019}. The discovery of these materials, together with Bi$_2$O$_2$Se, suggests that more highly polarizable semiconductors may be designed and synthesized in the complex chalcogenide family~\cite{Jaramillo2019}, many of which could also be in proximity to strain-induced paraelectric-ferroelectric phase transitions. It will be particularly interesting to see if exceptionally high low-temperature electron mobilities, as well as strain-induced ferroelectric transitions and mobility enhancements, could be observed in other highly polarizable chalcogenide semiconductors as well. Indeed, a high electron mobility at low temperature is needed for observing many quantum coherent phenomena such the quantum Hall effect and fractional quantum Hall effect~\cite{Tsui1982}, which could be useful for developing quantum devices and quantum technologies. The robust protection of electron mobility from ionized impurity scattering at room temperature, as well as the possibility of the coexistence of high electron mobility and strain-controllable ferroelectricity, could also make these materials attractive for fabricating novel electronic and neuromorphic devices.

\section{Conclusions}
In conclusion, our comprehensive study of the electron-phonon and electron-ionized impurity interactions in \BOS\  unravels the origin of the superior electron transport properties of this layered semiconductor, revealing a close connection between the extraordinary electron mobility of \BOS\ and a unique interlayer ferroelectric transition. We further demonstrate that the divergent increase of permittivity under a small biaxial strain in \BOS\ can be utilized to induce a giant enhancement of the low-temperature electron mobility and a robust protection of the room-temperature mobility, a phenomenon that can be generalized to other semiconductors in the vicinity of a ferroelectric transition. These results point to a promising new avenue of harnessing low-dimensional phase transitions~\cite{Li2021} for the discovery of layered semiconductors with unprecedented transport properties. 

\section*{Acknowledgements}

W.L. gratefully acknowledges the support by NSFC under Project No. 62004172, Westlake Multidisciplinary Research Initiative Center (MRIC) under Award No. 20200101, and the Westlake University HPC Center. X.L. thanks the support of NSFC under Project No. 11904294. The authors thank Prof. X-R. Zheng and Dr. C.-M. Dai for helpful discussions. W.L. would like to thank Prof. Boris Shklovskii for pointing out the existence of a lower bound in the Coulomb cross section of ionized impurity scattering.

\vspace{1cm}

\noindent \textsf{Supporting Information}: Method, including additional details on the computation of phonon-limited carrier mobility, ionized impurity scattering, and discussion on other possible carrier scattering mechanisms; origin of the large static dielectric permittivity of Bi$_2$O$_2$Se; additional details on the electronic and lattice dynamical properties of Bi$_2$O$_2$Se.

%\bibliography{reference}
%apsrev4-2.bst 2019-01-14 (MD) hand-edited version of apsrev4-1.bst
%Control: key (0)
%Control: author (8) initials jnrlst
%Control: editor formatted (1) identically to author
%Control: production of article title (0) allowed
%Control: page (0) single
%Control: year (1) truncated
%Control: production of eprint (0) enabled
%

\end{document}